\documentclass[showpacs,preprint,superscriptaddress,groupedaddress,longbibliography]{revtex4-1}  % for review and submission
\usepackage{xcolor}
\usepackage{graphicx}  % needed for figures
\usepackage{dcolumn}   % needed for some tables 
\usepackage{amssymb,amsmath,bm}  % for math
\usepackage{lipsum}  % for dummy text 
\def\psddlwt{-1.7} % wake temporal
\def\psddlws{-3}  % wake spatial
\def\psddldt{-2.7} % dns temporal
\def\psddlds{-2}

\begin{document}
% The following information is for internal review, please remove them for submission
%\widetext \leftline{\footnotesize Version 12. Last modified \today. \em  WORKING MANUSCRIPT -- NOT FOR PUBLIC DISTRIBUTION}

%=========================
%   TITLE:
%=========================
\title{Characterizing Elastic Turbulence in Channel Flows at Low Reynolds Number}
\author{Boyang Qin}  
\author{Paulo E. Arratia} 
\affiliation{\small
Department of Mechanical Engineering \& Applied Mechanics, University of Pennsylvania, Philadelphia, PA 19104, USA
}  
 
%=========================
%   ABSTRACT: 
%=========================
% abstract of no more than 600 characters, including spaces, which should be self-contained (no footnotes) for use in abstracting journals and % databases. Comments and Replies should not include an abstract.
\begin{abstract}
We experimentally investigate the flow of a viscoelastic fluid in a parallel shear geometry at low Reynolds number. As the flow becomes unstable via a nonlinear subcritical instability, velocimetry measurements show non-periodic fluctuations over a broad range of frequencies and wavelengths, consistent with the main features of elastic turbulence. Using the same experimental setup, we compare these features to those in the flow around cylinders, which is upstream to the parallel shear region; we find significant differences in power spectra scaling, intermittency statistics, and flow structures. We propose a simple mechanism to explain the growth of velocity fluctuations in parallel shear flows based on polymer stretching due to fluctuations in streamwise velocity gradients. 
\end{abstract}
%\keywords{Elastic turbulence, polymer physics, flow instability}
\maketitle \thispagestyle{plain}% turn on the page number
\maketitle \thispagestyle{plain}% turn on the page number
%=========================
%   INTRODUCTION, BKG
%=========================
%\section{Introduction}%

Unlike water, the flow of viscoelastic fluids such as polymeric and surfactant solutions can exhibit flow instabilities even in the absence of inertia, i.e. low Reynolds number (Re)~\cite{1989Muller,1990Larson,1993McKinley,1998Groisman,2002Arora,2006Arratia,2007Poole,2013Pan}. At high flow rates, flows of viscoelastic fluids exhibit a completely new type of chaotic behavior -- elastic turbulence -- that has no analogues in Newtonian liquids~\cite{2004Groisman, 2000Groisman,2001Groisman, 2010Fardin}. Purely elastic instabilities are found in many practical flows and understanding these instabilities is fundamental to our knowledge of how biological fluids (e.g. blood, vesicles, mucus) flow~\cite{2015Graham,2014Chaouqi,2014Steinberg,2008Muller}, chemical and polymer industries where flow instabilities have been plaguing processing for years~\cite{2003Bertola, 2004Denn}, and micro- and nano-fluidics where purely elastic instabilities were proposed as a way of effective mixing at small length scales~\cite{2001Groisman,2012Lindner,2009Lam,2014Clemens}.

These flow instabilities result from the development of polymeric elastic stresses in the fluid due to flow-induced changes in polymer conformation in solution. These stresses are strain-dependent, anisotropic, and depend on the nature of the flow \cite{1999LarsonBook}. Elastic stresses are often observed in systems where the mean flow has sufficient curvature, such as the flow between rotating disk~\cite{2000Groisman,2007Burghelea,2009Jun}, between concentric cylinders~\cite{1989Muller,1990Larson,1998Groisman, 2010Fardin}, curved channels~\cite{2001Groisman}, and around obstacles~\cite{2002Arora,2013Grilli}. In these systems, high velocity gradients and curved streamlines can stretch the polymer molecules, inducing elastic stress and flow instabilities~\cite{1999LarsonBook}. In fact, it has been argued that curvature is a necessary condition for infinitesimal perturbations to be amplified by the normal stress imbalances in viscoelastic flows \cite{1996McKinley,1996Pakdel,1996Shaqfeh}, and much of recent work on elastic turbulence has been devoted to geometries with curvature \cite{2013Grilli,2004Groisman}. 

Recent theoretical investigations, however, have shown that viscoelastic flows can be nonlinearly unstable even in parallel shear flows such as in straight pipes and channels at low Re~\cite{2004Meulenbroek, 2005Morozov, 2007Morozov, 2010Jovanovic,2011Jovanovic}. For example, nonlinear perturbation analysis~\cite{2004Meulenbroek, 2005Morozov, 2007Morozov} predicts a subcritical bifurcation from stable base states, while non-modal stability analysis predicts transient growth of perturbation~\cite{2010Jovanovic,2011Jovanovic}. Subsequent experiments in small pipes~\cite{2011Bonn} found unusually large velocity fluctuations that are activated at many time scales, but the subcritical nature of the instability was not established and no hysteretic behavior (a characteristic of subcritical instabilities) was reported. More recently, the existence of a nonlinear subcritical instability of viscoelastic fluid in a (micro)channel flow was reported in experiments~\cite{2013Pan}. It is shown that, in the absence of inertia (i.e. low Reynolds number), a finite level of perturbation is required to destabilize the flow and the resultant flow fluctuation is hysteretic~\cite{2013Pan}. This subcritical transition in viscoelastic channel flows is hence akin to the transition from laminar to turbulent flows of simple Newtonian fluids (e.g. water) in pipes, except that the governing parameter is the Weissenberg number (Wi), defined as the product of the fluid relaxation time $\lambda$ and the flow shear-rate $\dot{\gamma}$. However, the main features of the resulting unstable flow have yet to be well characterized and as a result, the flow of viscoelastic fluids in straight channel remains poorly understood.

\begin{figure}[h!]
\includegraphics[width=0.5\textwidth]{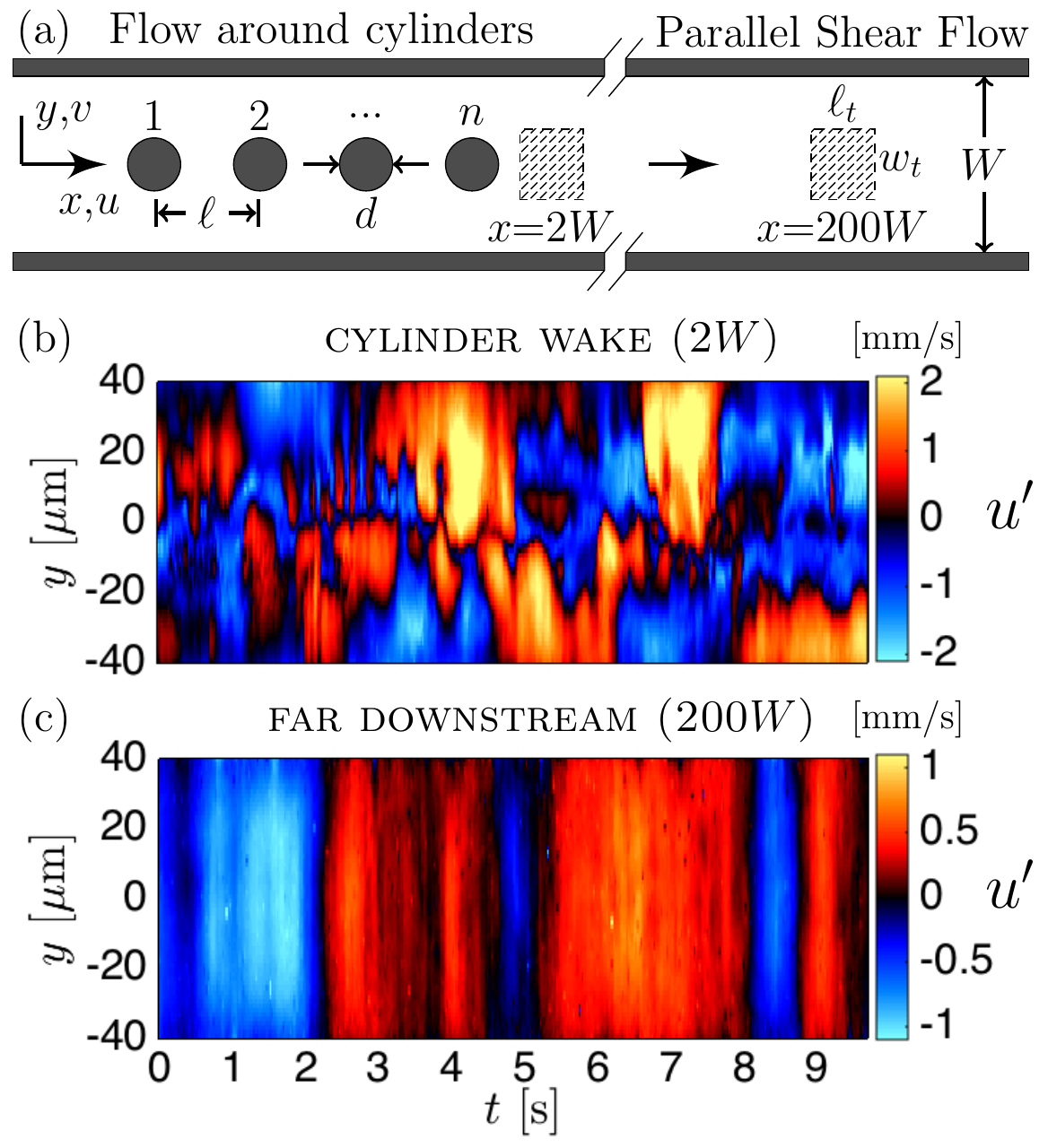}
\caption{\label{Fig1}(color online). (a) Schematic of the experimental channel geometry. % Boger fluids are passed through a square fluidic channel of width $W$=100$\mu$m, which will be used to non-dimensionalize lengths. Flow perturbation is induced by a spatially periodic array of $n$ cylinders, with diameter $d$=0.5 and streamwise spacing $\ell$=2. The subsequent parallel shear flow lasts for $\sim$300$W$. Time series of centerline velocity is monitored at various positions downstream of the last cylinder, with a rate of 1.5kHz and a duration of 300s. The measurement window lies on the mid-section plane in $z$ and has a width of $w_t$ = 0.1 and length $\ell_t$=0.7, shown by the shaded area. Spatial velocity fields are obtained at similar positions with a much larger window size of $w_s$= 0.9, $\ell_s$=1.2 and a grid resolution of $\sim 1\mu$m. 
% Spatial velocity fields are monitored at 2$W$ and 200$W$ downstream of the last cylinder and at the mid-section in $z$. A window size of width $w_s = 0.9$ and length $\ell_s=1.2$ is used, shown by the shaded area. Time series ($\sim$300 sec) of pointwise velocity is recorded at similar positions with a window size of $w_t = 0.1$ and length $\ell_t=0.7$. 
(b) Space-time plot of the streamwise velocity fluctuation $u^\prime$, immediately after the last cylinder, $x=2W$ and Wi$=10$. %High intensity fluctuations take the form of ``stripes'' of a wide range of sizes and are aperiodic in time. 
(c) The fluctuation landscape far downstream at 200$W$. % Here, we observe aperiodic ``bursts'' of various durations in time.
}
\end{figure}
\par
In this manuscript, we investigate the flow of a polymeric fluid in a straight micro-channel at low Re using particle tracking methods. The flow is excited using a linear array of cylinders and is monitored (i) immediately after the array of cylinders and (ii) far downstream. We find that both the flow next to the cylinder and that far downstream show features of elastic turbulence including velocity fluctuations excited over a broad range temporal frequencies and spatial length scales. There are, however, significant differences between those flows including the flow structure (c.f. Fig. 1b,c), velocity time series statistics, and temporal and spatial spectra decay. A simple mechanism is proposed for the sustained fluctuations observed in the parallel shear flow region based on a self-sustaining mechanism of energy feedback between the flow fluctuation and polymer elastic energy.%
\par
%\section{Methods}
The flow of a dilute polymeric solution is investigated using a straight microchannel with a square cross-section ($W=D=100 \,\mu$m). The microchannel is made of polydimethylsiloxane (PDMS) using standard soft-lithography methods.  The length of the microchannel is much larger than its width $L/W=330$, and it is partitioned into two regions. The first region is comprised of a linear array of cylinders that extends for $30W$. A total of 15 cylinders ($n=15$) are used in the linear array; a schematic is shown in Fig. 1(a). Each cylinder has a diameter $d$ of 0.5$W$ and is evenly spaced with a separation of $\ell=2W$; the last cylinder is at position $x=0$. The second region is a long parallel shear flow, which follows the initial linear array of cylinders and is $300W$ in length. More details on the channel design can be found elsewhere~\cite{2013Pan}.%
\par
The polymeric solution is prepared by adding 300 ppm of polyacrylamide (PAA, 18$\times10^6$ MW) in a viscous Newtonian solvent (90\% by weight glycerol aqueous solution). This solution possesses a nearly constant viscosity of approximately $\eta=200$ mPa$\cdot$s; for more information on the rheological properties of the fluid, see supplemental material (SM) and ~\cite{2013Pan}. A Newtonian solution, 90\% by weight glycerol in water, is also used for comparison. The Reynolds number (Re) is kept below 0.01, where Re $=\rho UH/\eta$, $U$ is the mean centerline velocity, $H$ is the channel half width, and $\rho$ is the fluid density. The strength of the elastic stresses compared to viscous stresses is characterized by the Weissenberg number~\cite{1993Magda,1993McKinley}, here defined as Wi($\dot\gamma$) $=N_1(\dot\gamma) /2\dot\gamma\eta(\dot\gamma)$, where $\dot\gamma= U/H$ is the shear rate and $N_1$ is the first normal stress difference. The fluid relaxation time is obtained from shear rheology data (see SM) and is defined as $\lambda(\dot\gamma) = N_1(\dot\gamma) /2\dot\gamma^2\eta$; values of $\lambda$ range from 0.1 to 1.0 seconds for the typical shear rates in the channel experiment. For the experiments presented here, the Weissenberg number is kept constant at approximately 10 and the number of cylinders at 15.  We note that the critical value of Wi for the onset of the \textit{subcritical instability} in the parallel flow region is Wi$_c=5.2$ for the type of disturbances (15 cylinders) introduced here~\cite{2013Pan}.

\par
The flow is characterized using particle tracking velocimetry. Fluorescent particles (0.6~$\mu$m in diameter) are dispersed in the fluids and imaged using an epifluorescent microscope and a high speed CMOS camera (up to $10^4$ fps). Spatially-resolved velocity fields are obtained by tracking particles in a rectangular window (width=0.9$W$, length=1.2$W$, centered at $y=0$) with a grid resolution of $\sim1~\mu$m. The resultant time resolution is $\Delta t=25$ ms. However, we can increase the time resolution and duration of the velocimetry measurements by decreasing the window size (width=0.1$W$, length=0.7$W$). This time-resolved measurement produces velocity time series with high resolution ($\Delta t=1$ ms) and relatively long sampling duration (up to 300 s).
\par
%
% \section{Results}
%========================
%   RESULTS & DISCUSSIONS:
%========================
We begin our flow analysis by measuring the flow streamwise velocity $u(x,y,t)$ in the wake of the last cylinder ($x=2W$) as well as in the parallel shear region ($x=200W$) using the spatially-resolved measurement (i.e. large window size). The streamwise velocity fluctuation $u'$ is obtained by subtracting the ensemble average $\langle u\rangle$ from the measured signal, $u'=u- \langle u\rangle$. Figure 1(b) shows the space time plot of $u'(y,t)$ along a cutline in the wall-normal direction ($y$-axis) at the cylinder wake region ($x=2W$ in Fig. 1a) of the channel. Here the spatial coordinate used is the wall-normal $y$ coordinate and the channel centerline is at $y=0$. The data show relatively large velocity fluctuation in the cylinder wake, with the amplitude reaching approximately 2 mm/s or 28\% of the overall channel centerline mean speed ($\sim7$~mm/s). Along the $y$ direction, we find that high intensity fluctuations are concentrated in the form of ``spots'', which are manifestations of streamwise streaks of high and low local velocity fluctuations. These streaks have a wide range of temporal durations and spatial sizes, as large as the cylinder diameter ($\sim50\,\mu$m) and as small as the velocity grid spacing ($\sim1\,\mu$m). Far downstream (200$W$, Fig. 1c), however, the flow is significantly different from that in the cylinder wake. We find that velocity fluctuations at $200W$ exist in the form of aperiodic ``bursts'' of various durations and appear to be spatially smoother in the wall-normal direction. We note that no appreciable fluctuations are found in the Newtonian case under similar conditions. Overall, we find markedly different flow structures as the fluid moves from regions near the cylinder (curved flows) to the parallel shear region.
\par
To quantify the temporal dynamics of the (unstable) flow, we measure the centerline velocity fluctuations $u_c'(t)$ for both Newtonian and polymeric solutions in the wake of the cylinder (Fig. 2a) and in the parallel shear region (Fig. 2b) using the small interrogation window. The data show significant velocity fluctuations for the viscoelastic fluid; the standard deviation (i.e. fluctuations) reaches approximately 10\% of the centerline mean, in both regions of the flow. No significant fluctuations are found in the Newtonian fluid case, shown in gray, under same conditions (i.e. flow rates). At both locations, the velocity fluctuations of the polymeric solution show an irregular pattern without an apparent periodicity, and the amplitudes of the centerline velocity variations are quite similar.  There are, however, differences between the flow in the wake of the cylinder (2$W$) and in the parallel shear region (200$W$). Specifically, the data show that far downstream (Fig. 2b), the velocity fluctuations in the high frequency range are weaker compared to those in the cylinder wake (Fig. 2a), as will be discussed below. In addition, the mean of $u_c'(t)$ at the cylinder wake is negatively biased towards the low velocity values. Physically, this means the flow at $2W$ is characterized by intermittent jumps to high velocities amidst dwelling at lower velocities, while far downstream (200$W$) the flow seems to fluctuate around the mean evenly.
\begin{figure}[h]
\centering
\includegraphics[width=0.7\textwidth]{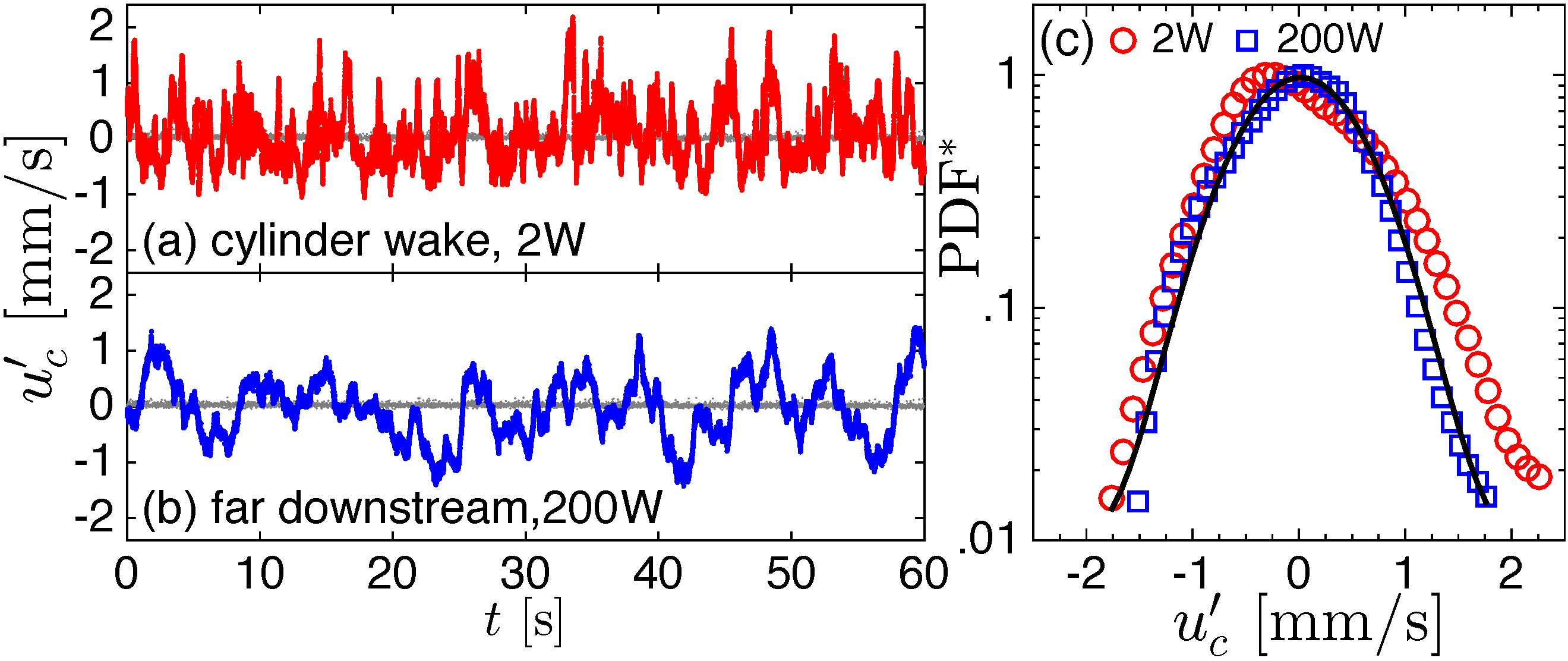}
\caption{\label{Fig2} (color online). Time series and the associated probability distribution of centerline velocity fluctuations $u_c'$ for $n=15$ and Wi$=10$ (a) Velocity records measured in the cylinder wake ($x=2W$). An interval of 60~s is shown out of the total duration of 300~s
% Aperiodic and high-frequency variations are observed, {\color{blue} compared to the steady flow of Newtonian solvent at identical flow rate, shown in gray, where the experimental noise level is 0.6\% of the centerline velocity}.  
(b) Velocity records measured far downstream in the parallel shear flow region (200$W$)
% One sees that far downstream the amplitude of the fluctuation is conserved, yet the high frequency variations are diminished compared to that in the cylinder wake. Moreover, in the immediate wake, the flow is marked by aperiodic spikes of high velocities amidst dwellings at low velocities, while 200W downstream, the flow fluctuates without significant bias. This observation is better quantified by 
(c) Probability distribution of the associated time series, normalized by the maximum of the probability density. Each curve includes 1.3$\times 10^6$ samples.
% Note the positive skewness in the distribution at the wake 2$W$, compared to the symmetric Gaussian distribution downstream at 200$W$, with curve representing Gaussian fit: the moment coefficient of skewness is 0.41 at $2W$ versus -0.07 at 200$W$ downstream. Such aperiodic spikes to higher velocity from a lower level right after the cylinder, are indicative of the sudden releases of polymer elastic energy into the flow as kinetic energy. The dynamics of the flow downstream, however, is marked by an even chance of low velocity versus high velocity, indicating a energy transfer back and forth between the polymer and the flow. This idea is further developed below by monitoring the fluctuation of the spatial velocity gradients, the kinematic signatures of polymer stretching.
}
\end{figure}
\par
The contrast between the flow in these two locations can be further quantified by computing the normalized probability distribution of $u_c'(t)$ (Fig. 2c). In the cylinder wake, we find that the mode of the distribution has a negative bias towards lower velocity, consistent with the data in Fig. 2(a). We also find a pronounced tail towards high velocities, which indicates that the distribution is positively skewed. In the parallel shear region (Fig. 2c), by contrast, we find a symmetric distribution that is well represented by a Gaussian fit (solid line). Consequently, the skewness of the distribution is 0.41 at $2W$, compared to the much lower 0.06 at 200$W$. {\color{black} We believe that near the cylinder (2$W$), the observed aperiodic jumps in $u_c'(t)$ are associated with the sudden release of elastic energy by polymer molecules into the flow (analogous to the intermittently injection of elastic energy in \cite{2006Burghelea,2009Jun}). Far downstream (200$W$), on the other hand, the even likelihood of velocity above and below the mean value indicates a unbiased energy transfer back and forth between the polymer and the flow. This idea is further developed below by monitoring the fluctuations of the spatial velocity gradients, the random components of the flow that drive the stretching of polymers~\cite{2010Liu}.}

% of a $u_c'$ in the  wake of the last curved cylinder at 2$W$. The flow is strongly perturbed ($n$=15) and energetic (Wi=10). Large centerline velocity fluctuation that is highly aperiodic and fast-varying is observed. In this case, the centerline mean $\langle u_c\rangle$ is about 3mm/s at 2$W$ and 7mm/s at 200$W$ downstream. Note that (c), the Newtonian solvent at the same location, perturbation level and flow rate displays no fluctuation.  (b) The measurement far downstream at 200$W$ at identical flow conditions. One sees that the characteristic time scale of the fluctuation is larger compared to that in the cylinder wake. Further, the amplitude of the fluctuation is preserved after being convected 200$W$ down the parallel shear flow region. This is shown in panel (d), the normalized probability distribution of velocity fluctuations for the cases in question (each 1.3 million samples). Lines represent Gaussian fit. While the distribution far downstream at 200W is Gaussian, the distribution in the immediate cylinder wake deviates from Gaussian. A heavy bias towards the positive end and an exponential tail are observed. [2006 Steinberg] has observed a similar bias but towards the negative end, in the input power distribution for the elastic turbulence between two counter rotating plates.
%
%
\par
Next, we analyze the velocity fluctuations by computing the frequency power spectra. Figure 3(a) shows the power spectra of the centerline velocity for $n=15$ and Wi$=10$, both polymeric and Newtonian solutions. The data show that the viscoelastic fluid flow is excited at a broad range of frequencies $f$ at all measured channel locations (from $2W$ to $200W$). This feature is one of the main hallmarks of elastic turbulence, which is most often observed in curved geometries~\cite{2000Groisman}.
\par
\begin{figure}[h]
\includegraphics[width=0.8\textwidth]{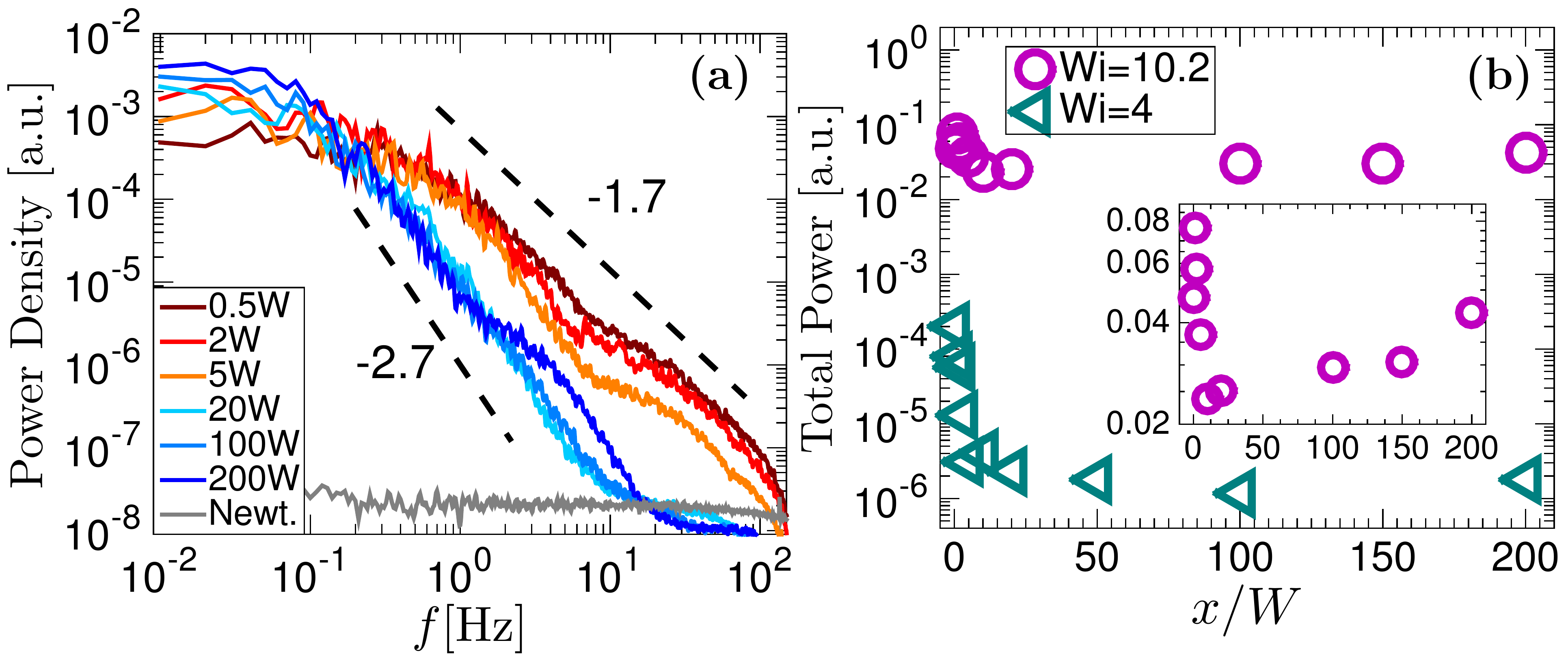}
\caption{\label{Fig3} (color online). Development of the frequency power spectra and total power of the centerline velocity along the channel. (a) frequency spectra, at positions immediately around the curved cylinder to far down the parallel shear region, Wi$=10$, $n=15$. (b) total spectral power contained in the velocity fluctuation, summed from the dominant range 0.01 to 100Hz. Inset is a zoom-in of the Wi$=10$ case.
}
\end{figure}
\par
Figure 3(a) also shows a gradual decay of the frequency power spectrum, following $f^{\psddlwt}$ in the wake of the cylinder ($2W$). {\color{black} We note that this value ($\psddlwt$) is significantly higher than the value of $-3.4$ reported in a recent 2D simulation of an Oldroyd-B fluid flowing in a channel with periodic array of cylinders~\cite{2013Grilli}. Moreover, the power law exponent $\psddlwt$ is relatively high compared to experiments of viscoelastic flows in closed systems with curved streamlines; for example, a $-3.3$ exponent was reported in a serpentine flow~\cite{2001Groisman}. However, Groisman and Steinberg~\cite{2004Groisman} reported a high power decay exponent, between -1.1 and -2.2, in experiments in a Taylor-Couette geometry, where rotation period is close the polymer relaxation time.} We note that in our flow geometry, the time scale associate with the flow round the linear array of cylinders ranges from $U/n\ell\sim$1 Hz to $U/d\sim$100 Hz and represents the frequency by which the mean flow is perturbed by the periodic cylinder array. This range overlaps with the polymer relaxation time scales (1-10 Hz) over a frequency decade. We believe that the abnormally gradual decay of the power spectra in the immediate wake of the cylinders is likely the result of the overlap of these two time scales.
\par
As the flow moves downstream from the array of cylinders into the parallel shear flow region, however, we observe clear developments in the frequency spectra. We find that, in a few channel widths after the last cylinder, the energy decreases in the high frequency range (10-100 Hz), which corresponds to the periodic perturbation introduced by the cylinders. At $x=20W$, the decrease in high frequency fluctuations intensifies across two frequency decades. On the other hand, the power in the low frequency range (0.01-0.1 Hz) of the spectrum increases. The combined result is that, after 20$W$, velocity fluctuations are increasingly dominated by low frequency variations and the power law decay becomes steeper, following $f^{\psddldt}$. This behavior is different from that reported by ~\cite{2011Bonn}, where spectral power is reduced across all frequencies as a function of distance from the entrance and the decay law ($-1.5$) was nearly the same as the flow moves downstream. By contrast, we find that the decay exponent changes from $-1.7$ near the cylinders to $-2.7$ far downstream and that the energy contained in the high frequency range near the cylinders seems to shift toward the low frequency range in the parallel shear region.
\par
Next, we compute the total spectral power by summing over the dominant frequency range (0.01-100 Hz). This is equivalent to the standard deviation of the time series if all valid frequencies are used. We perform this analysis for flows above and below the onset of the subcritical instability in the parallel shear region (Wi$_c = 5$ for $n=$15). Figure 3(b) show the evolution of the total energy down the channel for Wi=4 ($<$Wi$_c$) and Wi=10 ($>$Wi$_c$); $n=$ 15 for all flows investigated here. For the Wi=4 case, where the flow is not energetic enough to trigger velocity fluctuations downstream in the parallel shear flow, we find a sharp decay of total power by two orders of magnitude. The Wi=10 case sees a initial decay in total power within the first 20$W$. However, after $x=20W$, the trend reverses and follows a steady increase downstream into the parallel shear flow region (Fig. 3b inset), despite the dissipative environment (Re $\sim$0.01). Such persistence of fluctuation energy suggests a self-sustaining mechanism that we try to elucidate below. We note that $20W$ corresponds roughly to $\sim4\lambda U$ for the Wi=10 flow, where $U$ is the centerline mean velocity.

% One naturally raises the question of whether this is a dissipative effect or a transfer of energy from one end of the spectrum to the other.  Indeed, for the strongly energetic case of Wi=10.2, the total power is in fact conserved along channel length. Under the strong viscous dissipative environment(Re$\sim$0.01), such persistence of fluctuation energy suggests some self-sustaining mechanism. On the other hand, for flow of much lower Wi (e.g. Wi=4).
%
\begin{figure}[h]
\centering
\includegraphics[width=0.96\textwidth]{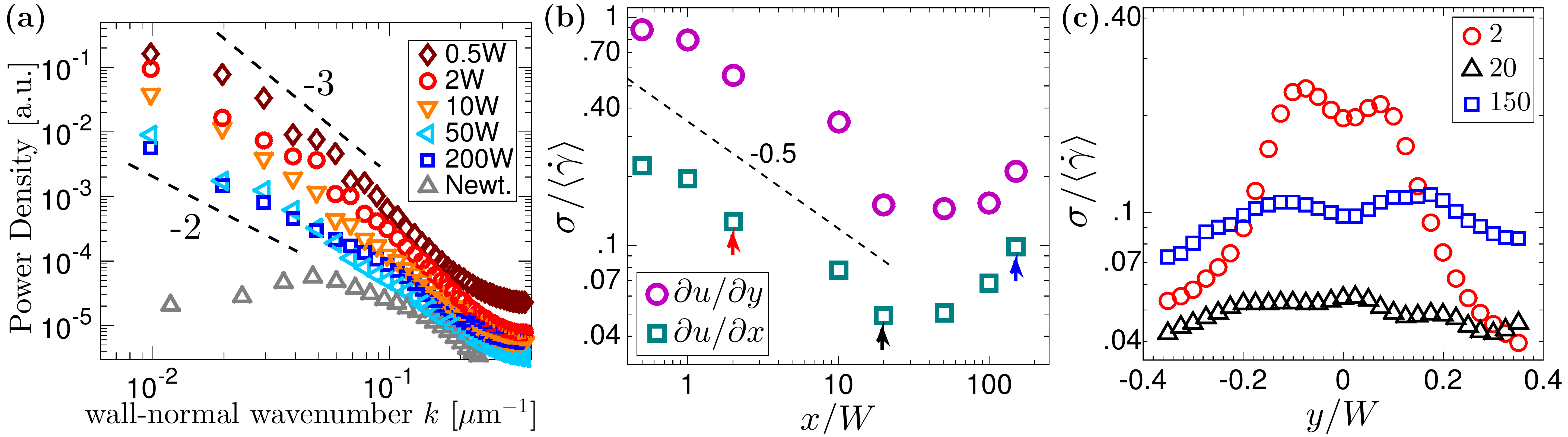}
\caption{\label{Fig4} (color online). Spatial characteristics of the (unstable) flow evolution along the channel for Wi$=10, n=15$. (a) Spatial power spectra as a function of wall-normal (y-axis, see Fig. 1b,c) wavenumber $k$ (spatial frequency) of the velocity fluctuation fields at various channel positions. (b) The rms variation $\sigma$ of shear ${\partial u}/{\partial y}$ and elongational  ${\partial u}/{\partial x}$ components of the velocity gradient, normalized by the spatial mean shear rate $\langle \dot\gamma \rangle$ in the parallel shear flow. (c) Elongation component of the rms profile across channel width $y$ immediately in the cylinder wake (2$W$), at the end of the cylinder flow decay (20$W$), and far downstream in the parallel shear flow region (150$W$). 
}
\end{figure}%
\par
We now turn our attention to the spatial features of the viscoelastic flow. Figure 4(a) shows the spatial spectra of the velocity fluctuations $u'(x,y,t)$ in the wall-normal ($y$-axis) direction. This direction is orthogonal to the streamwise flow and provides insights into the flow structures in the direction of gradients in shear. We find that the flow is activated at a wide range of spatial length scales $l$ (from 100 $\mu$m down to approximately 5 $\mu$m) in the wall-normal direction and that spatial variations in $u'$ are much stronger near the array of cylinders than in the parallel shear region; see also Fig. 1(b,c). The spatial spectrum of the viscoelastic flow near the cylinder follows a $k^{\psddlws}$ decay.  Note that the wavenumber $k$ is here defined as $1/l$, which is the spatial frequency. As the fluid travels downstream into the parallel shear flow, the spatial fluctuations weaken and the spectrum follows $k^{\psddlds}$; the data also shows that these spatial fluctuations are almost uniform across the channel (see Fig. 1c). These results, along with the data shown in Fig. 3(a), indicate that even though the flow near the cylinders and in the parallel shear region possess features of elastic turbulence they are quite different in their structure. This shows the difference between elastic turbulence in flows in curved geometries and in straight channels in a single system.

%Yet the question remains as to how the velocity fluctuation persist and grow after 20$W$ in the parallel shear flow.
% Spatial power spectrum of the velocity fluctuation fields. (a) 2$W$ in the cylinder wake. Compared to the Newtonian noise and the Boger case without perturbation ($n=0$), the strongly perturbed viscoelastic flow at high flow rate Wi=10.2 is activated at a wide range of length scales. A spatial decay law of -3 is observed, indifferent to the level of perturbations $n$ and energy level Wi (add data). (b) spatial spectrum 200$W$ downstream. The spatial power density of the fluctuation is weaker compared to that in the wake of the cylinder and displays a decay law of -2.3. Here the fluctuation intensity depends on perturbation level (increasing $n$), consistent with the subcritical bifurcation reported by\cite{2013Pan}.
\par

So far we have shown that the flow of a polymeric fluid in a parallel shear geometry can sustain relatively large velocity fluctuations in both space and time even at low Re. These velocity fluctuations, far downstream from the initial perturbation, are most likely driven by the stretching of polymer molecules in the flow. To test this hypothesis, we measure the root mean square (rms) variation of the shearing (${\partial u}/{\partial y}$) and elongational (${\partial u}/{\partial x}$) components of the velocity gradient;  here the rms of quantity $A$ is defined as $\sigma = \langle (A - \langle A \rangle)^2\rangle^{1/2} $, similar to \cite{2010Liu}. These components (quantities) are known to mediate polymer stretching in random flows~\cite{2000Balkovsky,2000Chertkov,2008Gerashchenko,2010Liu}.
%So far we have found distinct features of the flow of a polymeric solution in the parallel shear region ($200W$), compared to the flow in the wake of the cylinders ($2W$), that include (i) the disappearance of high fluctuation spots or streaks (Fig. 1c), (ii) the diminished spatial variation in the shearing $y$ direction (Fig. 4a), and (iii) the increase in low frequency velocity variations in the streamwise direction (Fig. 3a). A possible mechanism that may explain these flow features is 

Figure 4(b) shows the rms variation $\sigma$ of ${\partial u}/{\partial y}$ and ${\partial u}/{\partial x}$ at various positions along the channel normalized by the spatial average shear rate $\langle \dot\gamma \rangle$ downstream in the parallel shear flow. Near the linear array of cylinders, we find significant fluctuations of the velocity gradients relative to the mean shear rate in the parallel shear flow. Moreover, the ${\partial u}/{\partial y}$ component dominates ${\partial u}/{\partial x}$ and both components decay as the polymeric solution flows downstream. These trends persist down to approximately $20W$ in the channel. However, at $x\gtrsim20W$, we find that both components of $\sigma/\langle\dot\gamma\rangle$ reverse trend and begin to increase as the fluid flows downstream. Concurrently, we observe that the fluctuation in the elongation component become increasingly comparable to that of the shear component. These trends clearly show a change in flow at or around $20W$ accompanied by an increase in velocity fluctuations and polymer stretching. This non-monotonic trend is also captured by plotting the spatial profile of $\sigma/\langle\dot\gamma\rangle$ for ${\partial u}/{\partial x}$ across the channel width ($y$-axis) for three different channel locations, shown in Fig. 4(c). The data suggest that polymer molecules are increasingly stretched by flow gradient in the streamwise direction beyond $20W$. 

To further demonstrate the magnitude of the fluctuation in velocity gradients is large enough to generate polymer stretching, we compute a Weissenberg number based on fluctuations in the velocity gradients. Here, Wi$_{rms}^s = \lambda(\dot\gamma) \cdot \sigma(\partial u/ \partial y)$ where the rms fluctuation of the shear gradient is non-dimensionalized by the polymer relaxation time. Using the relaxation time corresponding to Wi=10.2, we find that Wi$_{rms}^s\sim5.2$ in the cylinder wake ($x=2W$), while far downstream, it is approximately 2. Moreover, far downstream, the Weissenberg number based on the rms of elongational Wi$_{rms}^e \sim 1$. We note that the values of both Wi$_{rms}^s$ and Wi$_{rms}^e$ are near or larger than $1$, which suggest that the flow is able to generate sufficient polymer stretching \cite{1997Perkins,1999Smith}. 

%The spatial profile of this increase in elongation component is shown in the inset of Fig. 4(b), which shows that the critical position $20W$, elongational polymer stretching vanishes, yet downstream in the parallel shear flow, a uniform enhancement occurs across the width of the channel, consistent with streamwise features of space-time landscape of the velocity fluctuation in Fig.\ref{Fig1}(c). 
%
%The subcritical nature of the observed flow fluctuation \cite{2013Pan} could be viewed in a new way under the proposed mechanism. It is likely that a finite level of polymer stretching, as a result from the initial curvature upstream and flow rate, need to be reached before the self-sustaining mechanism between the energy migration between the flow and the polymer, can be initiated. 
%========================
%   CONCLUSIONS:
%========================
%\section{Conclusion}
\par
In summary, we investigate the flow of a viscoelastic fluid in a parallel shear geometry at low Re. This flow becomes unstable via a nonlinear subcritical instability above a critical Weissenberg number (Wi$_c=5.2$) if perturbations are strong enough~\cite{2013Pan}. Using spatially- and temporally-resolved velocimetry, we identified signatures of elastic turbulence in the parallel shear region. This flow contrasts in many ways with elastic turbulence near the array of cylinders, which we find in our previous experiments (same experimental setup) to be {\it linearly} unstable \cite{2013Pan}. Specifically, we find that the flow near cylinders is organized by streamwise streaks that manifest as ``spots'' in Fig. 1(b), while temporal burst that manifest as spanwise bands are found in the parallel shear region (Fig. 1c). Moreover, the energy contained in the high frequency range near the cylinders seems to shift toward the low frequency range in the parallel shear region (Fig. 3a). We provide a simple mechanism for the sustained (and even growth) of velocity fluctuations in the parallel shear region based on polymer stretching due to r.m.s. fluctuations of the velocity gradients in the streamwise ($x$-axis) direction. These results suggest the emergence of a new type of elastic turbulent state in parallel shear flows. 
\par
%========================
%   ACKOWNLEDGEMENT:
%========================
We would like to thank V. Steinberg, G. Voth, C. Wagner, A. Morozov, Y. Dubief, S. Kumar, and M. Jovanovic for fruitful discussions. This work was supported by NSF-CBET-1336171.
%
%
%========================
%   BIBLIO:
%========================
\bibliography{reference_v1}

\begin{thebibliography}{42}%
\makeatletter
\providecommand \@ifxundefined [1]{%
 \@ifx{#1\undefined}
}%
\providecommand \@ifnum [1]{%
 \ifnum #1\expandafter \@firstoftwo
 \else \expandafter \@secondoftwo
 \fi
}%
\providecommand \@ifx [1]{%
 \ifx #1\expandafter \@firstoftwo
 \else \expandafter \@secondoftwo
 \fi
}%
\providecommand \natexlab [1]{#1}%
\providecommand \enquote  [1]{``#1''}%
\providecommand \bibnamefont  [1]{#1}%
\providecommand \bibfnamefont [1]{#1}%
\providecommand \citenamefont [1]{#1}%
\providecommand \href@noop [0]{\@secondoftwo}%
\providecommand \href [0]{\begingroup \@sanitize@url \@href}%
\providecommand \@href[1]{\@@startlink{#1}\@@href}%
\providecommand \@@href[1]{\endgroup#1\@@endlink}%
\providecommand \@sanitize@url [0]{\catcode `\\12\catcode `\$12\catcode
  `\&12\catcode `\#12\catcode `\^12\catcode `\_12\catcode `\%12\relax}%
\providecommand \@@startlink[1]{}%
\providecommand \@@endlink[0]{}%
\providecommand \url  [0]{\begingroup\@sanitize@url \@url }%
\providecommand \@url [1]{\endgroup\@href {#1}{\urlprefix }}%
\providecommand \urlprefix  [0]{URL }%
\providecommand \Eprint [0]{\href }%
\providecommand \doibase [0]{http://dx.doi.org/}%
\providecommand \selectlanguage [0]{\@gobble}%
\providecommand \bibinfo  [0]{\@secondoftwo}%
\providecommand \bibfield  [0]{\@secondoftwo}%
\providecommand \translation [1]{[#1]}%
\providecommand \BibitemOpen [0]{}%
\providecommand \bibitemStop [0]{}%
\providecommand \bibitemNoStop [0]{.\EOS\space}%
\providecommand \EOS [0]{\spacefactor3000\relax}%
\providecommand \BibitemShut  [1]{\csname bibitem#1\endcsname}%
\let\auto@bib@innerbib\@empty
%</preamble>
\bibitem [{\citenamefont {Muller}\ \emph {et~al.}(1989)\citenamefont {Muller},
  \citenamefont {Larson},\ and\ \citenamefont {Shaqfeh}}]{1989Muller}%
  \BibitemOpen
  \bibfield  {author} {\bibinfo {author} {\bibfnamefont {S.J.}\ \bibnamefont
  {Muller}}, \bibinfo {author} {\bibfnamefont {R.G.}\ \bibnamefont {Larson}}, \
  and\ \bibinfo {author} {\bibfnamefont {E.S.G.}\ \bibnamefont {Shaqfeh}},\
  }\bibfield  {title} {\enquote {\bibinfo {title} {A purely elastic transition
  in taylor-couette flow},}\ }\href {\doibase 10.1007/BF01332920} {\bibfield
  {journal} {\bibinfo  {journal} {Rheol. Acta.}\ }\textbf {\bibinfo {volume}
  {28}},\ \bibinfo {pages} {499--503} (\bibinfo {year} {1989})}\BibitemShut
  {NoStop}%
\bibitem [{\citenamefont {Larson}\ \emph {et~al.}(1990)\citenamefont {Larson},
  \citenamefont {Shaqfeh},\ and\ \citenamefont {Muller}}]{1990Larson}%
  \BibitemOpen
  \bibfield  {author} {\bibinfo {author} {\bibfnamefont {R.~G.}\ \bibnamefont
  {Larson}}, \bibinfo {author} {\bibfnamefont {Eric S.~G.}\ \bibnamefont
  {Shaqfeh}}, \ and\ \bibinfo {author} {\bibfnamefont {S.~J.}\ \bibnamefont
  {Muller}},\ }\bibfield  {title} {\enquote {\bibinfo {title} {A purely elastic
  instability in taylor-couette flow},}\ }\href {\doibase
  10.1017/S0022112090001124} {\bibfield  {journal} {\bibinfo  {journal} {J.
  Fluid Mech.}\ }\textbf {\bibinfo {volume} {218}},\ \bibinfo {pages}
  {573--600} (\bibinfo {year} {1990})}\BibitemShut {NoStop}%
\bibitem [{\citenamefont {McKinley}\ \emph {et~al.}(1993)\citenamefont
  {McKinley}, \citenamefont {Armstrong},\ and\ \citenamefont
  {Brown}}]{1993McKinley}%
  \BibitemOpen
  \bibfield  {author} {\bibinfo {author} {\bibfnamefont {Gareth~H.}\
  \bibnamefont {McKinley}}, \bibinfo {author} {\bibfnamefont {Robert~C.}\
  \bibnamefont {Armstrong}}, \ and\ \bibinfo {author} {\bibfnamefont
  {Robert~A.}\ \bibnamefont {Brown}},\ }\bibfield  {title} {\enquote {\bibinfo
  {title} {The wake instability in viscoelastic flow past confined circular
  cylinders},}\ }\href {\doibase 10.1098/rsta.1993.0091} {\bibfield  {journal}
  {\bibinfo  {journal} {Proc. R. Soc. A}\ }\textbf {\bibinfo {volume} {344}},\
  \bibinfo {pages} {265--304} (\bibinfo {year} {1993})}\BibitemShut {NoStop}%
\bibitem [{\citenamefont {Groisman}\ and\ \citenamefont
  {Steinberg}(1998)}]{1998Groisman}%
  \BibitemOpen
  \bibfield  {author} {\bibinfo {author} {\bibfnamefont {Alexander}\
  \bibnamefont {Groisman}}\ and\ \bibinfo {author} {\bibfnamefont {Victor}\
  \bibnamefont {Steinberg}},\ }\bibfield  {title} {\enquote {\bibinfo {title}
  {Mechanism of elastic instability in couette flow of polymer solutions:
  Experiment},}\ }\href {\doibase http://dx.doi.org/10.1063/1.869764}
  {\bibfield  {journal} {\bibinfo  {journal} {Phys. Fluids}\ }\textbf {\bibinfo
  {volume} {10}},\ \bibinfo {pages} {2451--2463} (\bibinfo {year}
  {1998})}\BibitemShut {NoStop}%
\bibitem [{\citenamefont {Arora}\ \emph {et~al.}(2002)\citenamefont {Arora},
  \citenamefont {Sureshkumar},\ and\ \citenamefont {Khomami}}]{2002Arora}%
  \BibitemOpen
  \bibfield  {author} {\bibinfo {author} {\bibfnamefont {K.}~\bibnamefont
  {Arora}}, \bibinfo {author} {\bibfnamefont {R.}~\bibnamefont {Sureshkumar}},
  \ and\ \bibinfo {author} {\bibfnamefont {B.}~\bibnamefont {Khomami}},\
  }\bibfield  {title} {\enquote {\bibinfo {title} {Experimental investigation
  of purely elastic instabilities in periodic flows},}\ }\href {\doibase
  http://dx.doi.org/10.1016/S0377-0257(02)00131-3} {\bibfield  {journal}
  {\bibinfo  {journal} {J. Non-Newton. Fluid Mech.}\ }\textbf {\bibinfo
  {volume} {108}},\ \bibinfo {pages} {209 -- 226} (\bibinfo {year}
  {2002})}\BibitemShut {NoStop}%
\bibitem [{\citenamefont {Arratia}\ \emph {et~al.}(2006)\citenamefont
  {Arratia}, \citenamefont {Thomas}, \citenamefont {Diorio},\ and\
  \citenamefont {Gollub}}]{2006Arratia}%
  \BibitemOpen
  \bibfield  {author} {\bibinfo {author} {\bibfnamefont {P.~E.}\ \bibnamefont
  {Arratia}}, \bibinfo {author} {\bibfnamefont {C.~C.}\ \bibnamefont {Thomas}},
  \bibinfo {author} {\bibfnamefont {J.}~\bibnamefont {Diorio}}, \ and\ \bibinfo
  {author} {\bibfnamefont {J.~P.}\ \bibnamefont {Gollub}},\ }\bibfield  {title}
  {\enquote {\bibinfo {title} {Elastic instabilities of polymer solutions in
  cross-channel flow},}\ }\href {\doibase 10.1103/PhysRevLett.96.144502}
  {\bibfield  {journal} {\bibinfo  {journal} {Phys. Rev. Lett.}\ }\textbf
  {\bibinfo {volume} {96}},\ \bibinfo {pages} {144502} (\bibinfo {year}
  {2006})}\BibitemShut {NoStop}%
\bibitem [{\citenamefont {Poole}\ \emph {et~al.}(2007)\citenamefont {Poole},
  \citenamefont {Alves},\ and\ \citenamefont {Oliveira}}]{2007Poole}%
  \BibitemOpen
  \bibfield  {author} {\bibinfo {author} {\bibfnamefont {R.~J.}\ \bibnamefont
  {Poole}}, \bibinfo {author} {\bibfnamefont {M.~A.}\ \bibnamefont {Alves}}, \
  and\ \bibinfo {author} {\bibfnamefont {P.~J.}\ \bibnamefont {Oliveira}},\
  }\bibfield  {title} {\enquote {\bibinfo {title} {Purely elastic flow
  asymmetries},}\ }\href {\doibase 10.1103/PhysRevLett.99.164503} {\bibfield
  {journal} {\bibinfo  {journal} {Phys. Rev. Lett.}\ }\textbf {\bibinfo
  {volume} {99}},\ \bibinfo {pages} {164503} (\bibinfo {year}
  {2007})}\BibitemShut {NoStop}%
\bibitem [{\citenamefont {Pan}\ \emph {et~al.}(2013)\citenamefont {Pan},
  \citenamefont {Morozov}, \citenamefont {Wagner},\ and\ \citenamefont
  {Arratia}}]{2013Pan}%
  \BibitemOpen
  \bibfield  {author} {\bibinfo {author} {\bibfnamefont {L.}~\bibnamefont
  {Pan}}, \bibinfo {author} {\bibfnamefont {A.}~\bibnamefont {Morozov}},
  \bibinfo {author} {\bibfnamefont {C.}~\bibnamefont {Wagner}}, \ and\ \bibinfo
  {author} {\bibfnamefont {P.~E.}\ \bibnamefont {Arratia}},\ }\bibfield
  {title} {\enquote {\bibinfo {title} {Nonlinear elastic instability in channel
  flows at low reynolds numbers},}\ }\href {\doibase
  10.1103/PhysRevLett.110.174502} {\bibfield  {journal} {\bibinfo  {journal}
  {Phys. Rev. Lett.}\ }\textbf {\bibinfo {volume} {110}},\ \bibinfo {pages}
  {174502} (\bibinfo {year} {2013})}\BibitemShut {NoStop}%
\bibitem [{\citenamefont {Groisman}\ and\ \citenamefont
  {Steinberg}(2004)}]{2004Groisman}%
  \BibitemOpen
  \bibfield  {author} {\bibinfo {author} {\bibfnamefont {Alexander}\
  \bibnamefont {Groisman}}\ and\ \bibinfo {author} {\bibfnamefont {Victor}\
  \bibnamefont {Steinberg}},\ }\bibfield  {title} {\enquote {\bibinfo {title}
  {Elastic turbulence in curvilinear flows of polymer solutions},}\ }\href
  {\doibase 10.1088/1367-2630/6/1/029} {\bibfield  {journal} {\bibinfo
  {journal} {New J. Phys.}\ }\textbf {\bibinfo {volume} {6}},\ \bibinfo {pages}
  {29} (\bibinfo {year} {2004})}\BibitemShut {NoStop}%
\bibitem [{\citenamefont {Groisman}\ and\ \citenamefont
  {Steinberg}(2000)}]{2000Groisman}%
  \BibitemOpen
  \bibfield  {author} {\bibinfo {author} {\bibfnamefont {Alexander}\
  \bibnamefont {Groisman}}\ and\ \bibinfo {author} {\bibfnamefont {Victor}\
  \bibnamefont {Steinberg}},\ }\bibfield  {title} {\enquote {\bibinfo {title}
  {Elastic turbulence in a polymer solution flow},}\ }\href {\doibase
  10.1038/35011019} {\bibfield  {journal} {\bibinfo  {journal} {Nature}\
  }\textbf {\bibinfo {volume} {405}},\ \bibinfo {pages} {53--55} (\bibinfo
  {year} {2000})}\BibitemShut {NoStop}%
\bibitem [{\citenamefont {Groisman}\ and\ \citenamefont
  {Steinberg}(2001)}]{2001Groisman}%
  \BibitemOpen
  \bibfield  {author} {\bibinfo {author} {\bibfnamefont {Alexander}\
  \bibnamefont {Groisman}}\ and\ \bibinfo {author} {\bibfnamefont {Victor}\
  \bibnamefont {Steinberg}},\ }\bibfield  {title} {\enquote {\bibinfo {title}
  {Efficient mixing at low reynolds numbers using polymer additives},}\ }\href
  {\doibase 10.1038/35073524} {\bibfield  {journal} {\bibinfo  {journal}
  {Nature}\ }\textbf {\bibinfo {volume} {410}},\ \bibinfo {pages} {905--908}
  (\bibinfo {year} {2001})}\BibitemShut {NoStop}%
\bibitem [{\citenamefont {Fardin}\ \emph {et~al.}(2010)\citenamefont {Fardin},
  \citenamefont {Lopez}, \citenamefont {Croso}, \citenamefont {Gr\'egoire},
  \citenamefont {Cardoso}, \citenamefont {McKinley},\ and\ \citenamefont
  {Lerouge}}]{2010Fardin}%
  \BibitemOpen
  \bibfield  {author} {\bibinfo {author} {\bibfnamefont {M.~A.}\ \bibnamefont
  {Fardin}}, \bibinfo {author} {\bibfnamefont {D.}~\bibnamefont {Lopez}},
  \bibinfo {author} {\bibfnamefont {J.}~\bibnamefont {Croso}}, \bibinfo
  {author} {\bibfnamefont {G.}~\bibnamefont {Gr\'egoire}}, \bibinfo {author}
  {\bibfnamefont {O.}~\bibnamefont {Cardoso}}, \bibinfo {author} {\bibfnamefont
  {G.~H.}\ \bibnamefont {McKinley}}, \ and\ \bibinfo {author} {\bibfnamefont
  {S.}~\bibnamefont {Lerouge}},\ }\bibfield  {title} {\enquote {\bibinfo
  {title} {Elastic turbulence in shear banding wormlike micelles},}\ }\href
  {\doibase 10.1103/PhysRevLett.104.178303} {\bibfield  {journal} {\bibinfo
  {journal} {Phys. Rev. Lett.}\ }\textbf {\bibinfo {volume} {104}},\ \bibinfo
  {pages} {178303} (\bibinfo {year} {2010})}\BibitemShut {NoStop}%
\bibitem [{\citenamefont {Henr\'{\i}quez~Rivera}\ \emph
  {et~al.}(2015)\citenamefont {Henr\'{\i}quez~Rivera}, \citenamefont {Sinha},\
  and\ \citenamefont {Graham}}]{2015Graham}%
  \BibitemOpen
  \bibfield  {author} {\bibinfo {author} {\bibfnamefont {Rafael~G.}\
  \bibnamefont {Henr\'{\i}quez~Rivera}}, \bibinfo {author} {\bibfnamefont
  {Kushal}\ \bibnamefont {Sinha}}, \ and\ \bibinfo {author} {\bibfnamefont
  {Michael~D.}\ \bibnamefont {Graham}},\ }\bibfield  {title} {\enquote
  {\bibinfo {title} {Margination regimes and drainage transition in confined
  multicomponent suspensions},}\ }\href {\doibase
  10.1103/PhysRevLett.114.188101} {\bibfield  {journal} {\bibinfo  {journal}
  {Phys. Rev. Lett.}\ }\textbf {\bibinfo {volume} {114}},\ \bibinfo {pages}
  {188101} (\bibinfo {year} {2015})}\BibitemShut {NoStop}%
\bibitem [{\citenamefont {Thi\'ebaud}\ \emph {et~al.}(2014)\citenamefont
  {Thi\'ebaud}, \citenamefont {Shen}, \citenamefont {Harting},\ and\
  \citenamefont {Misbah}}]{2014Chaouqi}%
  \BibitemOpen
  \bibfield  {author} {\bibinfo {author} {\bibfnamefont {Marine}\ \bibnamefont
  {Thi\'ebaud}}, \bibinfo {author} {\bibfnamefont {Zaiyi}\ \bibnamefont
  {Shen}}, \bibinfo {author} {\bibfnamefont {Jens}\ \bibnamefont {Harting}}, \
  and\ \bibinfo {author} {\bibfnamefont {Chaouqi}\ \bibnamefont {Misbah}},\
  }\bibfield  {title} {\enquote {\bibinfo {title} {Prediction of anomalous
  blood viscosity in confined shear flow},}\ }\href {\doibase
  10.1103/PhysRevLett.112.238304} {\bibfield  {journal} {\bibinfo  {journal}
  {Phys. Rev. Lett.}\ }\textbf {\bibinfo {volume} {112}},\ \bibinfo {pages}
  {238304} (\bibinfo {year} {2014})}\BibitemShut {NoStop}%
\bibitem [{\citenamefont {Levant}\ and\ \citenamefont
  {Steinberg}(2014)}]{2014Steinberg}%
  \BibitemOpen
  \bibfield  {author} {\bibinfo {author} {\bibfnamefont {Michael}\ \bibnamefont
  {Levant}}\ and\ \bibinfo {author} {\bibfnamefont {Victor}\ \bibnamefont
  {Steinberg}},\ }\bibfield  {title} {\enquote {\bibinfo {title} {Complex
  dynamics of compound vesicles in linear flow},}\ }\href {\doibase
  10.1103/PhysRevLett.112.138106} {\bibfield  {journal} {\bibinfo  {journal}
  {Phys. Rev. Lett.}\ }\textbf {\bibinfo {volume} {112}},\ \bibinfo {pages}
  {138106} (\bibinfo {year} {2014})}\BibitemShut {NoStop}%
\bibitem [{\citenamefont {Gulati}\ \emph {et~al.}(2008)\citenamefont {Gulati},
  \citenamefont {Liepmann},\ and\ \citenamefont {Muller}}]{2008Muller}%
  \BibitemOpen
  \bibfield  {author} {\bibinfo {author} {\bibfnamefont {Shelly}\ \bibnamefont
  {Gulati}}, \bibinfo {author} {\bibfnamefont {Dorian}\ \bibnamefont
  {Liepmann}}, \ and\ \bibinfo {author} {\bibfnamefont {Susan~J.}\ \bibnamefont
  {Muller}},\ }\bibfield  {title} {\enquote {\bibinfo {title} {{Elastic
  secondary flows of semidilute DNA solutions in abrupt 90 degrees
  microbends}},}\ }\href@noop {} {\bibfield  {journal} {\bibinfo  {journal}
  {Phys. Rev. E}\ }\textbf {\bibinfo {volume} {78}},\ \bibinfo {pages} {036314}
  (\bibinfo {year} {2008})}\BibitemShut {NoStop}%
\bibitem [{\citenamefont {Meulenbroek}\ \emph {et~al.}(2003)\citenamefont
  {Meulenbroek}, \citenamefont {Storm}, \citenamefont {Bertola}, \citenamefont
  {Wagner}, \citenamefont {Bonn},\ and\ \citenamefont {van
  Saarloos}}]{2003Bertola}%
  \BibitemOpen
  \bibfield  {author} {\bibinfo {author} {\bibfnamefont {Bernard}\ \bibnamefont
  {Meulenbroek}}, \bibinfo {author} {\bibfnamefont {Cornelis}\ \bibnamefont
  {Storm}}, \bibinfo {author} {\bibfnamefont {Volfango}\ \bibnamefont
  {Bertola}}, \bibinfo {author} {\bibfnamefont {Christian}\ \bibnamefont
  {Wagner}}, \bibinfo {author} {\bibfnamefont {Daniel}\ \bibnamefont {Bonn}}, \
  and\ \bibinfo {author} {\bibfnamefont {Wim}\ \bibnamefont {van Saarloos}},\
  }\bibfield  {title} {\enquote {\bibinfo {title} {Intrinsic route to melt
  fracture in polymer extrusion: A weakly nonlinear subcritical instability of
  viscoelastic poiseuille flow},}\ }\href {\doibase
  10.1103/PhysRevLett.90.024502} {\bibfield  {journal} {\bibinfo  {journal}
  {Phys. Rev. Lett.}\ }\textbf {\bibinfo {volume} {90}},\ \bibinfo {pages}
  {024502} (\bibinfo {year} {2003})}\BibitemShut {NoStop}%
\bibitem [{\citenamefont {Denn}(2004)}]{2004Denn}%
  \BibitemOpen
  \bibfield  {author} {\bibinfo {author} {\bibfnamefont {MM}~\bibnamefont
  {Denn}},\ }\bibfield  {title} {\enquote {\bibinfo {title} {Fifty years of
  non-newtonian fluid dynamics},}\ }\href@noop {} {\bibfield  {journal}
  {\bibinfo  {journal} {AIChE J.}\ }\textbf {\bibinfo {volume} {50}},\ \bibinfo
  {pages} {2335--2345} (\bibinfo {year} {2004})}\BibitemShut {NoStop}%
\bibitem [{\citenamefont {Zilz}\ \emph {et~al.}(2012)\citenamefont {Zilz},
  \citenamefont {Poole}, \citenamefont {Alves}, \citenamefont {Bartolo},
  \citenamefont {Levach{\'e}},\ and\ \citenamefont {Lindner}}]{2012Lindner}%
  \BibitemOpen
  \bibfield  {author} {\bibinfo {author} {\bibfnamefont {J}~\bibnamefont
  {Zilz}}, \bibinfo {author} {\bibfnamefont {RJ}~\bibnamefont {Poole}},
  \bibinfo {author} {\bibfnamefont {MA}~\bibnamefont {Alves}}, \bibinfo
  {author} {\bibfnamefont {D}~\bibnamefont {Bartolo}}, \bibinfo {author}
  {\bibfnamefont {B}~\bibnamefont {Levach{\'e}}}, \ and\ \bibinfo {author}
  {\bibfnamefont {A}~\bibnamefont {Lindner}},\ }\bibfield  {title} {\enquote
  {\bibinfo {title} {Geometric scaling of a purely elastic flow instability in
  serpentine channels},}\ }\href@noop {} {\bibfield  {journal} {\bibinfo
  {journal} {J. Fluid Mech.}\ }\textbf {\bibinfo {volume} {712}},\ \bibinfo
  {pages} {203--218} (\bibinfo {year} {2012})}\BibitemShut {NoStop}%
\bibitem [{\citenamefont {Lam}\ \emph {et~al.}(2009)\citenamefont {Lam},
  \citenamefont {Gan}, \citenamefont {Nguyen},\ and\ \citenamefont
  {Lie}}]{2009Lam}%
  \BibitemOpen
  \bibfield  {author} {\bibinfo {author} {\bibfnamefont {YC}~\bibnamefont
  {Lam}}, \bibinfo {author} {\bibfnamefont {HY}~\bibnamefont {Gan}}, \bibinfo
  {author} {\bibfnamefont {Nam-Trung}\ \bibnamefont {Nguyen}}, \ and\ \bibinfo
  {author} {\bibfnamefont {H}~\bibnamefont {Lie}},\ }\bibfield  {title}
  {\enquote {\bibinfo {title} {Micromixer based on viscoelastic flow
  instability at low reynolds number},}\ }\href@noop {} {\bibfield  {journal}
  {\bibinfo  {journal} {Biomicrofluidics}\ }\textbf {\bibinfo {volume} {3}},\
  \bibinfo {pages} {014106} (\bibinfo {year} {2009})}\BibitemShut {NoStop}%
\bibitem [{\citenamefont {Scholz}\ \emph {et~al.}(2014)\citenamefont {Scholz},
  \citenamefont {Wirner}, \citenamefont {Gomez-Solano},\ and\ \citenamefont
  {Bechinger}}]{2014Clemens}%
  \BibitemOpen
  \bibfield  {author} {\bibinfo {author} {\bibfnamefont {Christian}\
  \bibnamefont {Scholz}}, \bibinfo {author} {\bibfnamefont {Frank}\
  \bibnamefont {Wirner}}, \bibinfo {author} {\bibfnamefont {Juan~Ruben}\
  \bibnamefont {Gomez-Solano}}, \ and\ \bibinfo {author} {\bibfnamefont
  {Clemens}\ \bibnamefont {Bechinger}},\ }\bibfield  {title} {\enquote
  {\bibinfo {title} {Enhanced dispersion by elastic turbulence in porous
  media},}\ }\href@noop {} {\bibfield  {journal} {\bibinfo  {journal} {EPL}\
  }\textbf {\bibinfo {volume} {107}},\ \bibinfo {pages} {54003} (\bibinfo
  {year} {2014})}\BibitemShut {NoStop}%
\bibitem [{\citenamefont {Larson}(1999)}]{1999LarsonBook}%
  \BibitemOpen
  \bibfield  {author} {\bibinfo {author} {\bibfnamefont {Ronald~G}\
  \bibnamefont {Larson}},\ }\href@noop {} {\emph {\bibinfo {title} {The
  structure and rheology of complex fluids}}},\ Vol.~\bibinfo {volume} {33}\
  (\bibinfo  {publisher} {Oxford university press New York},\ \bibinfo {year}
  {1999})\BibitemShut {NoStop}%
\bibitem [{\citenamefont {Burghelea}\ \emph {et~al.}(2007)\citenamefont
  {Burghelea}, \citenamefont {Segre},\ and\ \citenamefont
  {Steinberg}}]{2007Burghelea}%
  \BibitemOpen
  \bibfield  {author} {\bibinfo {author} {\bibfnamefont {Teodor}\ \bibnamefont
  {Burghelea}}, \bibinfo {author} {\bibfnamefont {Enrico}\ \bibnamefont
  {Segre}}, \ and\ \bibinfo {author} {\bibfnamefont {Victor}\ \bibnamefont
  {Steinberg}},\ }\bibfield  {title} {\enquote {\bibinfo {title} {Elastic
  turbulence in von karman swirling flow between two disks},}\ }\href {\doibase
  10.1063/1.2732234} {\bibfield  {journal} {\bibinfo  {journal} {Phys. Fluids}\
  }\textbf {\bibinfo {volume} {19}},\ \bibinfo {pages} {053104} (\bibinfo
  {year} {2007})}\BibitemShut {NoStop}%
\bibitem [{\citenamefont {Jun}\ and\ \citenamefont
  {Steinberg}(2009)}]{2009Jun}%
  \BibitemOpen
  \bibfield  {author} {\bibinfo {author} {\bibfnamefont {Yonggun}\ \bibnamefont
  {Jun}}\ and\ \bibinfo {author} {\bibfnamefont {Victor}\ \bibnamefont
  {Steinberg}},\ }\bibfield  {title} {\enquote {\bibinfo {title} {Power and
  pressure fluctuations in elastic turbulence over a wide range of polymer
  concentrations},}\ }\href {\doibase 10.1103/PhysRevLett.102.124503}
  {\bibfield  {journal} {\bibinfo  {journal} {Phys. Rev. Lett.}\ }\textbf
  {\bibinfo {volume} {102}},\ \bibinfo {pages} {124503} (\bibinfo {year}
  {2009})}\BibitemShut {NoStop}%
\bibitem [{\citenamefont {Grilli}\ \emph {et~al.}(2013)\citenamefont {Grilli},
  \citenamefont {V\'azquez-Quesada},\ and\ \citenamefont
  {Ellero}}]{2013Grilli}%
  \BibitemOpen
  \bibfield  {author} {\bibinfo {author} {\bibfnamefont {Muzio}\ \bibnamefont
  {Grilli}}, \bibinfo {author} {\bibfnamefont {Adolfo}\ \bibnamefont
  {V\'azquez-Quesada}}, \ and\ \bibinfo {author} {\bibfnamefont {Marco}\
  \bibnamefont {Ellero}},\ }\bibfield  {title} {\enquote {\bibinfo {title}
  {Transition to turbulence and mixing in a viscoelastic fluid flowing inside a
  channel with a periodic array of cylindrical obstacles},}\ }\href {\doibase
  10.1103/PhysRevLett.110.174501} {\bibfield  {journal} {\bibinfo  {journal}
  {Phys. Rev. Lett.}\ }\textbf {\bibinfo {volume} {110}},\ \bibinfo {pages}
  {174501} (\bibinfo {year} {2013})}\BibitemShut {NoStop}%
\bibitem [{\citenamefont {McKinley}\ \emph {et~al.}(1996)\citenamefont
  {McKinley}, \citenamefont {Pakdel},\ and\ \citenamefont
  {Aztekin}}]{1996McKinley}%
  \BibitemOpen
  \bibfield  {author} {\bibinfo {author} {\bibfnamefont {Gareth~H.}\
  \bibnamefont {McKinley}}, \bibinfo {author} {\bibfnamefont {Peyman}\
  \bibnamefont {Pakdel}}, \ and\ \bibinfo {author} {\bibfnamefont {Alparslan}\
  \bibnamefont {Aztekin}},\ }\bibfield  {title} {\enquote {\bibinfo {title}
  {Rheological and geometric scaling of purely elastic flow instabilities},}\
  }\href {\doibase 10.1016/S0377-0257(96)01453-X} {\bibfield  {journal}
  {\bibinfo  {journal} {J. Non-Newton. Fluid Mech.}\ }\textbf {\bibinfo
  {volume} {67}},\ \bibinfo {pages} {19 -- 47} (\bibinfo {year}
  {1996})}\BibitemShut {NoStop}%
\bibitem [{\citenamefont {Pakdel}\ and\ \citenamefont
  {McKinley}(1996)}]{1996Pakdel}%
  \BibitemOpen
  \bibfield  {author} {\bibinfo {author} {\bibfnamefont {Peyman}\ \bibnamefont
  {Pakdel}}\ and\ \bibinfo {author} {\bibfnamefont {Gareth~H.}\ \bibnamefont
  {McKinley}},\ }\bibfield  {title} {\enquote {\bibinfo {title} {Elastic
  instability and curved streamlines},}\ }\href {\doibase
  10.1103/PhysRevLett.77.2459} {\bibfield  {journal} {\bibinfo  {journal}
  {Phys. Rev. Lett.}\ }\textbf {\bibinfo {volume} {77}},\ \bibinfo {pages}
  {2459--2462} (\bibinfo {year} {1996})}\BibitemShut {NoStop}%
\bibitem [{\citenamefont {Shaqfeh}(1996)}]{1996Shaqfeh}%
  \BibitemOpen
  \bibfield  {author} {\bibinfo {author} {\bibfnamefont {E~S~G}\ \bibnamefont
  {Shaqfeh}},\ }\bibfield  {title} {\enquote {\bibinfo {title} {Purely elastic
  instabilities in viscometric flows},}\ }\href {\doibase
  10.1146/annurev.fl.28.010196.001021} {\bibfield  {journal} {\bibinfo
  {journal} {Annu. Rev. Fluid Mech.}\ }\textbf {\bibinfo {volume} {28}},\
  \bibinfo {pages} {129--185} (\bibinfo {year} {1996})}\BibitemShut {NoStop}%
\bibitem [{\citenamefont {Meulenbroek}\ \emph {et~al.}(2004)\citenamefont
  {Meulenbroek}, \citenamefont {Storm}, \citenamefont {Morozov},\ and\
  \citenamefont {van Saarloos}}]{2004Meulenbroek}%
  \BibitemOpen
  \bibfield  {author} {\bibinfo {author} {\bibfnamefont {Bernard}\ \bibnamefont
  {Meulenbroek}}, \bibinfo {author} {\bibfnamefont {Cornelis}\ \bibnamefont
  {Storm}}, \bibinfo {author} {\bibfnamefont {Alexander~N.}\ \bibnamefont
  {Morozov}}, \ and\ \bibinfo {author} {\bibfnamefont {Wim}\ \bibnamefont {van
  Saarloos}},\ }\bibfield  {title} {\enquote {\bibinfo {title} {Weakly
  nonlinear subcritical instability of visco-elastic poiseuille flow},}\ }\href
  {\doibase 10.1016/j.jnnfm.2003.09.003} {\bibfield  {journal} {\bibinfo
  {journal} {J. Non-Newton. Fluid Mech.}\ }\textbf {\bibinfo {volume} {116}},\
  \bibinfo {pages} {235 -- 268} (\bibinfo {year} {2004})}\BibitemShut {NoStop}%
\bibitem [{\citenamefont {Morozov}\ and\ \citenamefont {van
  Saarloos}(2005)}]{2005Morozov}%
  \BibitemOpen
  \bibfield  {author} {\bibinfo {author} {\bibfnamefont {Alexander~N.}\
  \bibnamefont {Morozov}}\ and\ \bibinfo {author} {\bibfnamefont {Wim}\
  \bibnamefont {van Saarloos}},\ }\bibfield  {title} {\enquote {\bibinfo
  {title} {Subcritical finite-amplitude solutions for plane couette flow of
  viscoelastic fluids},}\ }\href {\doibase 10.1103/PhysRevLett.95.024501}
  {\bibfield  {journal} {\bibinfo  {journal} {Phys. Rev. Lett.}\ }\textbf
  {\bibinfo {volume} {95}},\ \bibinfo {pages} {024501} (\bibinfo {year}
  {2005})}\BibitemShut {NoStop}%
\bibitem [{\citenamefont {Morozov}\ and\ \citenamefont {van
  Saarloos}(2007)}]{2007Morozov}%
  \BibitemOpen
  \bibfield  {author} {\bibinfo {author} {\bibfnamefont {Alexander~N.}\
  \bibnamefont {Morozov}}\ and\ \bibinfo {author} {\bibfnamefont {Wim}\
  \bibnamefont {van Saarloos}},\ }\bibfield  {title} {\enquote {\bibinfo
  {title} {An introductory essay on subcritical instabilities and the
  transition to turbulence in visco-elastic parallel shear flows},}\ }\href
  {\doibase http://dx.doi.org/10.1016/j.physrep.2007.03.004} {\bibfield
  {journal} {\bibinfo  {journal} {Phys. Rep.}\ }\textbf {\bibinfo {volume}
  {447}},\ \bibinfo {pages} {112--143} (\bibinfo {year} {2007})}\BibitemShut
  {NoStop}%
\bibitem [{\citenamefont {Jovanovic}\ and\ \citenamefont
  {Kumar}(2010)}]{2010Jovanovic}%
  \BibitemOpen
  \bibfield  {author} {\bibinfo {author} {\bibfnamefont {Mihailo~R.}\
  \bibnamefont {Jovanovic}}\ and\ \bibinfo {author} {\bibfnamefont {Satish}\
  \bibnamefont {Kumar}},\ }\bibfield  {title} {\enquote {\bibinfo {title}
  {Transient growth without inertia},}\ }\href {\doibase 10.1063/1.3299324}
  {\bibfield  {journal} {\bibinfo  {journal} {Phys. Fluids}\ }\textbf {\bibinfo
  {volume} {22}},\ \bibinfo {pages} {023101} (\bibinfo {year}
  {2010})}\BibitemShut {NoStop}%
\bibitem [{\citenamefont {Jovanović}\ and\ \citenamefont
  {Kumar}(2011)}]{2011Jovanovic}%
  \BibitemOpen
  \bibfield  {author} {\bibinfo {author} {\bibfnamefont {Mihailo~R.}\
  \bibnamefont {Jovanović}}\ and\ \bibinfo {author} {\bibfnamefont {Satish}\
  \bibnamefont {Kumar}},\ }\bibfield  {title} {\enquote {\bibinfo {title}
  {Nonmodal amplification of stochastic disturbances in strongly elastic
  channel flows},}\ }\href {\doibase
  http://dx.doi.org/10.1016/j.jnnfm.2011.02.010} {\bibfield  {journal}
  {\bibinfo  {journal} {J. Non-Newton. Fluid Mech.}\ }\textbf {\bibinfo
  {volume} {166}},\ \bibinfo {pages} {755 -- 778} (\bibinfo {year}
  {2011})}\BibitemShut {NoStop}%
\bibitem [{\citenamefont {Bonn}\ \emph {et~al.}(2011)\citenamefont {Bonn},
  \citenamefont {Ingremeau}, \citenamefont {Amarouchene},\ and\ \citenamefont
  {Kellay}}]{2011Bonn}%
  \BibitemOpen
  \bibfield  {author} {\bibinfo {author} {\bibfnamefont {D.}~\bibnamefont
  {Bonn}}, \bibinfo {author} {\bibfnamefont {F.}~\bibnamefont {Ingremeau}},
  \bibinfo {author} {\bibfnamefont {Y.}~\bibnamefont {Amarouchene}}, \ and\
  \bibinfo {author} {\bibfnamefont {H.}~\bibnamefont {Kellay}},\ }\bibfield
  {title} {\enquote {\bibinfo {title} {Large velocity fluctuations in
  small-reynolds-number pipe flow of polymer solutions},}\ }\href {\doibase
  10.1103/PhysRevE.84.045301} {\bibfield  {journal} {\bibinfo  {journal} {Phys.
  Rev. E}\ }\textbf {\bibinfo {volume} {84}},\ \bibinfo {pages} {045301}
  (\bibinfo {year} {2011})}\BibitemShut {NoStop}%
\bibitem [{\citenamefont {Magda}\ \emph {et~al.}(1993)\citenamefont {Magda},
  \citenamefont {Lee}, \citenamefont {Muller},\ and\ \citenamefont
  {Larson}}]{1993Magda}%
  \BibitemOpen
  \bibfield  {author} {\bibinfo {author} {\bibfnamefont {Jules~J.}\
  \bibnamefont {Magda}}, \bibinfo {author} {\bibfnamefont {Chang~Soon}\
  \bibnamefont {Lee}}, \bibinfo {author} {\bibfnamefont {Susan~J.}\
  \bibnamefont {Muller}}, \ and\ \bibinfo {author} {\bibfnamefont {Ronald~G.}\
  \bibnamefont {Larson}},\ }\bibfield  {title} {\enquote {\bibinfo {title}
  {Rheology, flow instabilities, and shear-induced diffusion in polystyrene
  solutions},}\ }\href {\doibase 10.1021/ma00059a032} {\bibfield  {journal}
  {\bibinfo  {journal} {Macromolecules}\ }\textbf {\bibinfo {volume} {26}},\
  \bibinfo {pages} {1696--1706} (\bibinfo {year} {1993})}\BibitemShut {NoStop}%
\bibitem [{\citenamefont {Burghelea}\ \emph {et~al.}(2006)\citenamefont
  {Burghelea}, \citenamefont {Segre},\ and\ \citenamefont
  {Steinberg}}]{2006Burghelea}%
  \BibitemOpen
  \bibfield  {author} {\bibinfo {author} {\bibfnamefont {Teodor}\ \bibnamefont
  {Burghelea}}, \bibinfo {author} {\bibfnamefont {Enrico}\ \bibnamefont
  {Segre}}, \ and\ \bibinfo {author} {\bibfnamefont {Victor}\ \bibnamefont
  {Steinberg}},\ }\bibfield  {title} {\enquote {\bibinfo {title} {Role of
  elastic stress in statistical and scaling properties of elastic
  turbulence},}\ }\href {\doibase 10.1103/PhysRevLett.96.214502} {\bibfield
  {journal} {\bibinfo  {journal} {Phys. Rev. Lett.}\ }\textbf {\bibinfo
  {volume} {96}},\ \bibinfo {pages} {214502} (\bibinfo {year}
  {2006})}\BibitemShut {NoStop}%
\bibitem [{\citenamefont {Liu}\ and\ \citenamefont
  {Steinberg}(2010)}]{2010Liu}%
  \BibitemOpen
  \bibfield  {author} {\bibinfo {author} {\bibfnamefont {Y.}~\bibnamefont
  {Liu}}\ and\ \bibinfo {author} {\bibfnamefont {V.}~\bibnamefont
  {Steinberg}},\ }\bibfield  {title} {\enquote {\bibinfo {title} {Stretching of
  polymer in a random flow: Effect of a shear rate},}\ }\href
  {http://stacks.iop.org/0295-5075/90/i=4/a=44005} {\bibfield  {journal}
  {\bibinfo  {journal} {EPL}\ }\textbf {\bibinfo {volume} {90}},\ \bibinfo
  {pages} {44005} (\bibinfo {year} {2010})}\BibitemShut {NoStop}%
\bibitem [{\citenamefont {Balkovsky}\ \emph {et~al.}(2000)\citenamefont
  {Balkovsky}, \citenamefont {Fouxon},\ and\ \citenamefont
  {Lebedev}}]{2000Balkovsky}%
  \BibitemOpen
  \bibfield  {author} {\bibinfo {author} {\bibfnamefont {E.}~\bibnamefont
  {Balkovsky}}, \bibinfo {author} {\bibfnamefont {A.}~\bibnamefont {Fouxon}}, \
  and\ \bibinfo {author} {\bibfnamefont {V.}~\bibnamefont {Lebedev}},\
  }\bibfield  {title} {\enquote {\bibinfo {title} {Turbulent dynamics of
  polymer solutions},}\ }\href {\doibase 10.1103/PhysRevLett.84.4765}
  {\bibfield  {journal} {\bibinfo  {journal} {Phys. Rev. Lett.}\ }\textbf
  {\bibinfo {volume} {84}},\ \bibinfo {pages} {4765--4768} (\bibinfo {year}
  {2000})}\BibitemShut {NoStop}%
\bibitem [{\citenamefont {Chertkov}(2000)}]{2000Chertkov}%
  \BibitemOpen
  \bibfield  {author} {\bibinfo {author} {\bibfnamefont {Michael}\ \bibnamefont
  {Chertkov}},\ }\bibfield  {title} {\enquote {\bibinfo {title} {Polymer
  stretching by turbulence},}\ }\href {\doibase 10.1103/PhysRevLett.84.4761}
  {\bibfield  {journal} {\bibinfo  {journal} {Phys. Rev. Lett.}\ }\textbf
  {\bibinfo {volume} {84}},\ \bibinfo {pages} {4761--4764} (\bibinfo {year}
  {2000})}\BibitemShut {NoStop}%
\bibitem [{\citenamefont {Gerashchenko}\ and\ \citenamefont
  {Steinberg}(2008)}]{2008Gerashchenko}%
  \BibitemOpen
  \bibfield  {author} {\bibinfo {author} {\bibfnamefont {Sergiy}\ \bibnamefont
  {Gerashchenko}}\ and\ \bibinfo {author} {\bibfnamefont {Victor}\ \bibnamefont
  {Steinberg}},\ }\bibfield  {title} {\enquote {\bibinfo {title} {Critical
  slowing down in polymer dynamics near the coil-stretch transition in
  elongation flow},}\ }\href {\doibase 10.1103/PhysRevE.78.040801} {\bibfield
  {journal} {\bibinfo  {journal} {Phys. Rev. E}\ }\textbf {\bibinfo {volume}
  {78}},\ \bibinfo {pages} {040801} (\bibinfo {year} {2008})}\BibitemShut
  {NoStop}%
\bibitem [{\citenamefont {Perkins}\ \emph {et~al.}(1997)\citenamefont
  {Perkins}, \citenamefont {Smith},\ and\ \citenamefont {Chu}}]{1997Perkins}%
  \BibitemOpen
  \bibfield  {author} {\bibinfo {author} {\bibfnamefont {Thomas~T.}\
  \bibnamefont {Perkins}}, \bibinfo {author} {\bibfnamefont {Douglas~E.}\
  \bibnamefont {Smith}}, \ and\ \bibinfo {author} {\bibfnamefont {Steven}\
  \bibnamefont {Chu}},\ }\bibfield  {title} {\enquote {\bibinfo {title} {Single
  polymer dynamics in an elongational flow},}\ }\href {\doibase
  10.1126/science.276.5321.2016} {\bibfield  {journal} {\bibinfo  {journal}
  {Science}\ }\textbf {\bibinfo {volume} {276}},\ \bibinfo {pages} {2016--2021}
  (\bibinfo {year} {1997})}\BibitemShut {NoStop}%
\bibitem [{\citenamefont {Smith}\ \emph {et~al.}(1999)\citenamefont {Smith},
  \citenamefont {Babcock},\ and\ \citenamefont {Chu}}]{1999Smith}%
  \BibitemOpen
  \bibfield  {author} {\bibinfo {author} {\bibfnamefont {Douglas~E.}\
  \bibnamefont {Smith}}, \bibinfo {author} {\bibfnamefont {Hazen~P.}\
  \bibnamefont {Babcock}}, \ and\ \bibinfo {author} {\bibfnamefont {Steven}\
  \bibnamefont {Chu}},\ }\bibfield  {title} {\enquote {\bibinfo {title}
  {Single-polymer dynamics in steady shear flow},}\ }\href {\doibase
  10.1126/science.283.5408.1724} {\bibfield  {journal} {\bibinfo  {journal}
  {Science}\ }\textbf {\bibinfo {volume} {283}},\ \bibinfo {pages} {1724--1727}
  (\bibinfo {year} {1999})}\BibitemShut {NoStop}%
\end{thebibliography}%
%
%{ \color{blue} Significance:
%\begin{itemize}
%\item Establishing evidence for purely elastic turbulence in pipe flows, which has not be shown before.
%\item How does Elasto-Inertia turbulence in pipe behave in the low Re limit. Insights into the mechanism of turb. drag reduction.
%\item Energy dynamics/cascade in straight pipe flow of polymers after finite perturbation, in the low Re limit. How does it evolve downstream?
%\end{itemize}
%}
%
%{\color{orange}
%Elastic turbulence can exist in PSF. Moreover, it is a new type of turbulence. We present evidence showing that the dynamics of elastic turbulence in parallel shear flow is drastically different from that after the curved posts. Namely:
%\begin{enumerate}
%\item Curved geometry ET decays till 20W. PSF turbulence emerges after 20W.
%\item Velocity time series: pdf distribution reveals sustained energy flow back and forth between flow and polymer, in contrast to that in the cylinder wake.
%\item Space-time plot: distinct structures.
%\item Spectra: distinct temoral and spatial decay scaling.
%\item we propose a self-sustaining mechanism of energy flow back and forth between the mean flow and polymer elastic energy.
%\end{enumerate}
%}

\end{document}